# Supervision of the ATLAS High Level Trigger System

S. Wheeler *
*University of Alberta, Centre for Subatomic Research, Edmonton, Alberta T6G 2N5, Canada*

J. Flammer
*EP Division, CERN, CH-1211 Geneva 23, Switzerland*

C. Meessen, Z. Qian, F. Touchard
*Centre de Physique des Particules de Marseille, Case 907, 163 avenue de Luminy, F-13288 Marseille Cedex 9, France*

A. Negri
*Istituto di Fisica Nucleare, Universita di Pavia & I.N.F.N., Via A. Bassi, 6, 27100, Pavia, Italy*

H. Zobernig
*University of Wisconsin, Department of Physics, Madison, WI 53706, USA*

on behalf of the ATLAS High Level Trigger Group [1]

* Presenter at the conference

The ATLAS High Level Trigger (HLT) system provides software-based event selection after the initial LVL1 hardware trigger. It is composed of two stages, the LVL2 trigger and the Event Filter. The HLT is implemented as software tasks running on large processor farms. An essential part of the HLT is the supervision system, which is responsible for configuring, coordinating, controlling and monitoring the many hundreds of processes running in the HLT. A prototype implementation of the supervision system, using tools from the ATLAS Online Software system is presented. Results from scalability tests are also presented where the supervision system was shown to be capable of controlling over 1000 HLT processes running on 230 nodes.

## 1. INTRODUCTION

ATLAS is a general-purpose particle physics experiment, currently under construction at the Large Hadron Collider (LHC) at CERN. It has been designed to exploit the full physics potential of the LHC including searches for as yet unobserved phenomena such as the Higgs boson and super-symmetry.

The ATLAS Trigger and Data Acquisition (TDAQ) system will have to deal with extremely high data rates, due both to the high bunch crossing frequency at the LHC (40 MHz) and the large amount of data produced by the ATLAS detector itself (~1.6 Mbyte per event). The task of the TDAQ system is to select from this unprecedented amount of data the most interesting events and save them for later analysis at a rate of about 200 per second. ATLAS relies on a three-level trigger system to perform the selection: a very fast, hardware-based LVL1 trigger, followed by two software-based triggers, the LVL2 trigger which is located before the Event Builder and the Event Filter (EF), after the Event Builder, which perform increasingly fine-grained selection of events at lower rates. The LVL2 and EF comprise the ATLAS High level trigger (HLT) system. The software running in the HLT may be split into three main functional areas:

- Event selection software, i.e. the physics selection algorithms, which analyzes event data and produces a trigger decision, either rejecting or accepting the event
- Dataflow software, which is responsible for transferring the event data and trigger decisions to and from the physics selection algorithms
- Supervision software, which is responsible for all aspects of software task management and control in the HLT

This paper will describe a prototype HLT supervision system, which has been implemented and subsequent tests to demonstrate its scalability.

Mandates of the supervision include:
- Software configuration of the farms and of the HLT software processes
- Synchronizing the HLT processes with data-taking activities in the rest of the experiment
- Monitoring the status of the HLT processes e.g. checking that they are running and restarting crashed process

Currently farm management (monitoring of computer hardware, operating system management etc.) is not included in the supervision system mandate.

Both the LVL2 trigger and the EF are implemented as software processes running on large processor farms consisting of commodity components connected via high-speed Ethernet. In view of this the supervision requirements for the two systems are very similar and an integrated HLT supervision system has been developed.

In practice the processor farms are split into a number of sub-farms. This is done for reasons of practicality, making it inherently more scalable and more fault-tolerant by allowing easy reconfiguration in case of a failure within a particular sub-farm. In the case of the EF it makes sense to associate one sub-farm to each output of the event-building switch. It is intended that the LVL2 farms will also be split into sub-farms, however, the actual network layout is still being investigated in order that it can be optimized.

## 2. PROTOTYPE HLT SUPERVISION SYSTEM

A prototype HLT supervision system has been implemented using services from the ATLAS Online Software system (OnlineSW). The OnlineSW system [2] is the software used to configure, control and monitor the TDAQ system but excludes the processing and transportation of physics data. It is generic and does not contain any elements that are detector-specific allowing it to be used throughout the TDAQ system including the trigger farms. It has been successfully adapted for use in the HLT.

In the HLT supervision prototype developed for use in the scalability tests the following tools from the OnlineSW were used to implement the mandates of the HLT supervision system listed in the previous section.

- Configuration databases [3] to describe which trigger processes run on which hosts within the farm
- Run Control system [4] to synchronize and coordinate data-taking activities within the HLT with the rest of the experiment
- DSA_Supervisor [4] to start, stop and monitor trigger processes.

The uses of these tools are described in more detail in the following sections.

### 2.1. Configuration Database

The configuration databases are implemented as xml files. For the scalability tests, each configuration had to be described by a number of xml configuration files. The trigger farms contain a very large number of nodes, each with very similar parameter, for example, the number of processes running selection algorithms per processing node. A program has been written in Tcl/Tk, which allows the necessary configuration database files be generated and subsequently modified, quickly and efficiently. The graphical user interface (GUI) from the program is shown in Figure 1. The GUI is simple to use and hides the complexity of the underlying OnlineSW configuration database. One practical issue is that the numbering scheme of any large-scale cluster is unlikely to be uniform due to problems with specific nodes at any one time. The GUI allows node numbers to be included in a particular sub-farm, either individually or as ranges, thereby allowing gaps in the numbering scheme to specified with the minimum of effort.

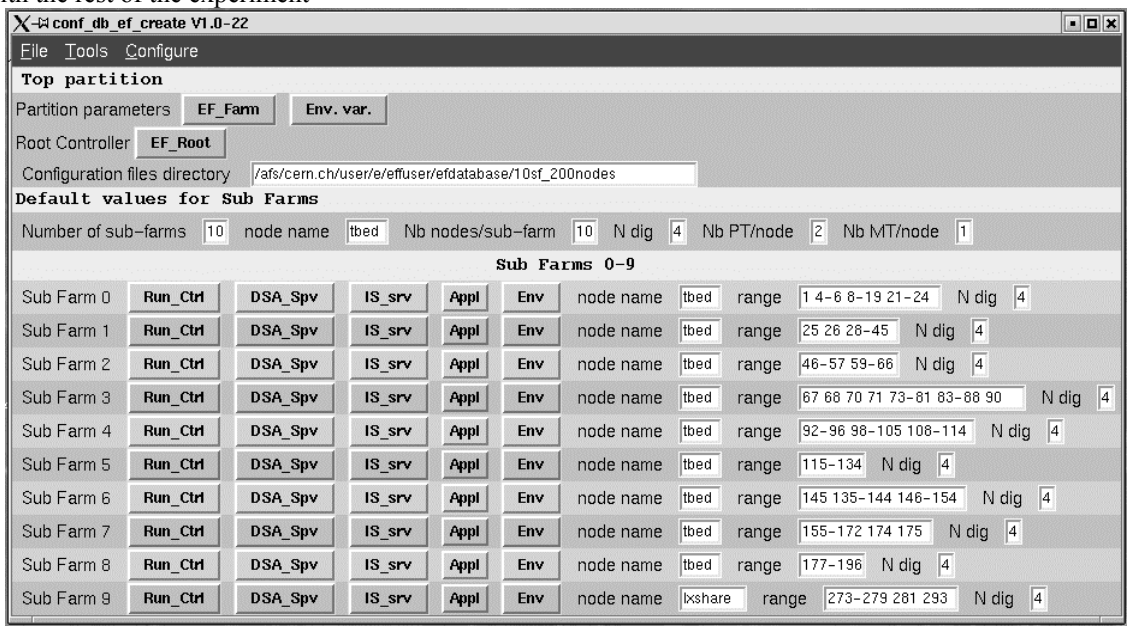

Figure 1: Graphical User Interface for database entry program

### 2.2. Run Control

Synchronization of the trigger processes with the rest of the TDAQ is achieved using the Run Control component. Run controllers, based on finite-state machines, are arranged in a hierarchical tree with one Run controller per sub-farm and one top-level farm controller. Commands from the operator are sent to the top-level farm Run controller, which forwards them to the sub-farm Run controllers. The sub-farm Run controllers try to change state by performing whatever action they need to do, for example starting all the trigger processes on a sub-farm. The sub-farm Run controllers inform the top-level farm Run controller if they have successfully completed the transition. Once all the sub-farm Run controllers have changed state the top-level controller can change state, hence synchronization is achieved across the whole farm. The sub-farm Run controllers are customizable. They are based on the OnlineSW Run control skeleton, which implements only the finite-state machine. The functions, which are performed at each transition of the finite-state machine, are implemented by the TDAQ system implementing the Run Control. In this case they are implemented to perform the actions required by the HLT system.

### 2.3. DSA_Supervisor

The DSA_Supervisor is used to provide process management and control. In the current version of the OnlineSW only one DSA_Supervisor process would normally be running in the system and would be responsible for starting all the processes described in the configuration databases. This does not scale well. However, the OnlineSW does not put any limitations on the number of DSA_Supervisors, which can run in the system. The prototype therefore uses the global DSA_Supervisor to manage the overall supervision infrastructure in conjunction with a dedicated DSA_Supervisor per sub-farm, which works in collaboration with the sub-farm Run controller to control and monitor the trigger processes in each sub-farm. This creates a much more scalable system and complies, at least in part with the proposed future design of the OnlineSW/HLT interface described in [5].

### 2.4. Controlling an HLT Farm

Figure 2 shows the sequence of operations used to prepare an HLT farm for data-taking activities. Referring to Figure 2a, an operator using the OnlineSW play_daq script [6], starts the OnlineSW infrastructure including amongst other things the Integrated Graphical User Interface (IGUI) [7], from which subsequent commands can be issued, and the global DSA_Supervisor process.

The "boot" command is then issued from the GUI (Figure 2b) to the DSA_Supervisor. On receiving this command the DSA_Supervisor reads the configuration database and starts the processes comprising the supervision infrastructure on the control hosts, i.e. a farm Run controller and a Run controller/DSA_Supervisor pair per sub-farm, on the correct processing nodes.

Once the infrastructure has been booted it is possible to send run control commands from the IGUI (Figure 2c). On receiving the "load" command from the GUI the farm Run controller forwards it to all the sub-farm Run controllers. In turn they ask the sub-farm DSA_Supervisors to start all the trigger processes on the sub-farm processing hosts. The DSA_Supervisors read the configuration database to determine on which processing nodes to start the trigger processes and starts them. Once all the processes are started the sub-farm DSA_Supervisors inform the sub-farm Run controllers, which are then able to complete their "load" transition.

Subsequent commands (Figure 2d) from the central console are directed to the trigger processes via the run control tree to prepare them for receiving data. Data-taking in the farm is stopped and the farm shutdown by reversing the sequence of actions described above.

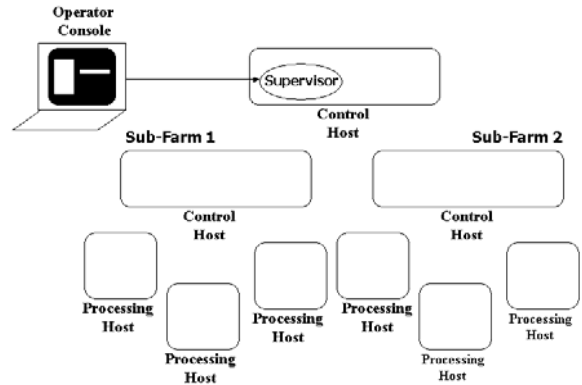

Figure 2a

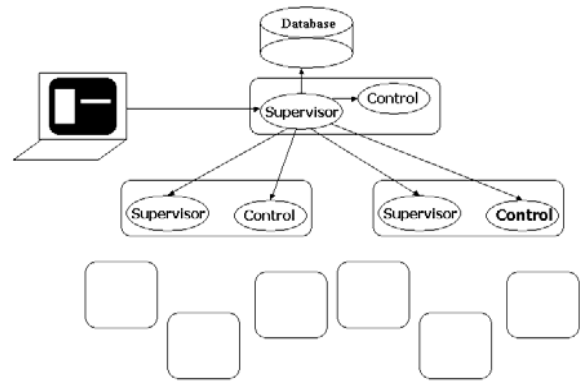

Figure 2b

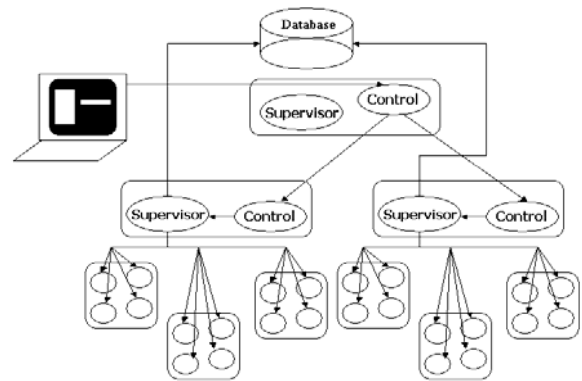

Figure 2c

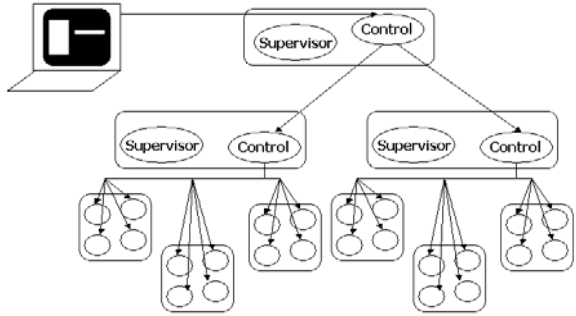

Figure 2d

## 2.5. Monitoring

Although not one of the features studied during the scalability tests, the supervision system is also responsible for carrying out monitoring activities. Monitoring has also been implemented using tools from the OnlineSW. The Information Service [8] is used for gathering statistical information. HLT processes write this information into information service servers for retrieval by others for display. The HLT processes can also send error and other informational messages to any other TDAQ component using the Message Reporting Service [9].

Figure 3: Example of the IGUI control panel showing

A panel to display the statistical information generated by HLT trigger processes is currently under development and can be integrated into the IGUI. An example of the panel in the IGUI showing information for an Event Filter Farm is shown in Figure 3. The left hand side of the IGUI is always visible and is used for issuing the DSA_Supervisor and Run control commands described in section 2.4. The right hand side displays one of a number of different panels according to the tab selected. The Event Filter panel, which reads and displays statistical information written by trigger tasks is shown. Summary sub-farm information is displayed in the top half of the panel. Clicking on a particular sub-farm displays detailed information in the panel below. Error and other informational messages sent via the Message Reporting system are displayed in a panel, which is always visible at the bottom of the IGUI.

## 3. SCALABILITY TESTS

A series of tests were carried out to determine the scalability of the control architecture described in section 2. The principal aim was to measure the time taken to perform the steps required to prepare various large configurations for data-taking (as described in section 2.4) and subsequently, the times taken to shut them down again. OnlineSW tools were used to make the timing measurements. Fault tolerance, error handling, reliability and monitoring were not within the scope of these tests. The tests were carried out on the IT LXPLUS Cluster at CERN. A total of 230 quad-processor nodes were available. Two types of configuration were studied:

- The number of processing nodes was kept constant but split into differing numbers of sub-farms.
- The number of sub-farms was kept constant however, the number of nodes per sub-farm were varied.

In the tests, the number of trigger processes running on each node was 4. Therefore, in the largest configurations studied, of the order of 1000 processes were under the control of the HLT prototype supervision system, representing approximately 10-20% of the HLT system, which will be used for the first ATLAS run.

## 4. RESULTS

Figure 4 shows the times to start and stop the top-level HLT supervision infrastructure for a constant number of processing nodes, but differing numbers of sub-farms. The line corresponding to "boot" is the time to start all the processes in the control infrastructure and the line corresponding to "shutdown" is the time taken to stop all the processes in the control infrastructure. An increase in times is seen with the number of controlled sub-farms. This is to be expected since the number of infrastructure process (Run Controller/DSA_Supervisor pairs) increases with the number of sub-farms. However, even for 21 sub-farms the times do not exceed 5 seconds.

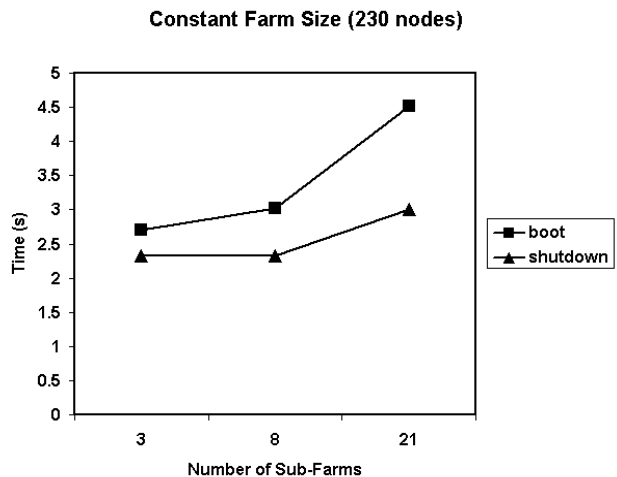

Figure 4: Times to start and stop HLT supervision infrastructure processes

Figure 5 shows the times required by the trigger processes to prepare for data-taking, i.e. it shows the times taken by the sub-farm Run Controllers to execute the various Run Control commands. As for the previous figure the graph shows results for a constant number of nodes, but differing numbers of sub-farms. In that the HLT supervision architecture is hierarchical, it allows preparation to occur in all sub-farms in parallel. Therefore, a decrease in times is seen with increasing numbers of sub-farms due to the smaller numbers of nodes per sub-farm. The "load" and "unload" lines indicate the times taken to start and stop all the trigger processes in the sub-farms. The "configure" line indicates the amount of time required by the sub-farm Run Controller to establish contact with the trigger processes it needs to control. The actions it performs during this time are reading the configuration database, creating communication objects, forwarding the command to all the controlled processes, in series and waiting for all their replies. The trigger processes themselves do not perform any action on receiving the configure command, they simply bounce it back to the Run Controller. The "start" and "stop" lines indicate the times taken to forward the command to all controlled processes which perform a simple action and reply. The "unconfigure" line indicates the time taken to forward the command and receive replies from the controlled processes (they do not perform any action) and then delete the communication objects in the run controller.

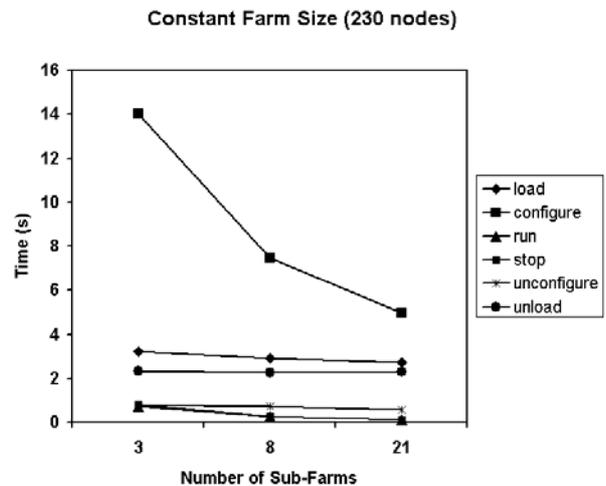

Figure 5: Times to prepare trigger processes for data-taking, as a function of number of sub-farms

The times taken for all configurations for all commands is again less than 5 seconds apart from the "configure" command. A bug was later discovered in the "configure" command in the sub-farm Run Controller. Subsequent tests following correction of the bug on a smaller testbed have shown these times should be about half of what is shown here. However, some non-linearity was still observed and needs to be understood. Possible explanations could be problems due to database access or inefficient setting up of the communication links with the controlled processes.

Figure 6 shows the times taken by the Run Control commands but for configurations in which the number of sub-farms was kept constant at 10 and the number of nodes per sub-farm varied. As expected an increase in times is seen as the number of nodes per sub-farm increases, due to the larger number of processes which need to be controlled. Again all commands take less than 5 seconds to complete apart from the "configure" command, possible explanations for which have already been given.

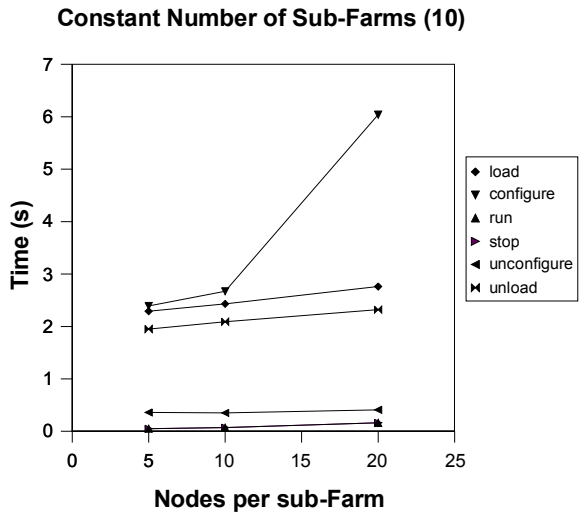

Figure 6: Times to prepare trigger processes for data-taking, as a function of number processing nodes per sub-farms

The statistical errors on the timing measurements are of the order of 0.1s.

## 5. CONCLUSIONS AND FUTURE

The results presented in the previous section are very promising for the implementation of the HLT supervision system for the first ATLAS run. All the operations required to startup, prepare for data-taking and shutdown large HLT farm configurations take of the order of a few seconds to complete. It was estimated that the largest configurations studied represent approximately 10-20% of the HLT system, which will be implemented for the first ATLAS run.

A number of enhancements of the HLT supervision system are foreseen. These will include, for example a combined Run control/DSA_Supervisor component and a distributed configuration database to match the distributed control structure. The communication between the sub-farm Run controllers and the trigger processes will be parallelized. Currently the Run controllers communicate with the trigger processes on a serial basis. This did not cause problems during the tests reported here, however, this would lead to delays if the trigger processes had lengthy actions to perform on particular Run Control commands.

## Acknowledgments

The authors wish to thank the ATLAS TDAQ Online Software group for their assistance during the tests. We would also like to thank the members of. the IT Division at CERN for providing access to the LXPLUS cluster for the scalability tests.

[a] S. Armstrong, J.T. Baines, C.P. Bee, M. Biglietti, M. Bosman, S. Brandt, A. Bogaerts, V. Boisvert, B. Caron, P. Casado, G. Cataldi,
D. Cavalli, M. Cervetto, G. Comune, A. Corso-Radu, A. Di Mattia, M. Diaz Gomez, A. dos Anjos, J. Drohan, N. Ellis, M. Elsing,
B. Epp, S. Falciano, A. Farilla, F. Etienne, S. George, V. Ghete, S. Gonzalez, M. Grothe, A. Kaczmarska, K. Karr, A. Khomich,
N. Konstantinidis, W. Krasny, W. Li, A. Lowe, L. Luminari, H. Ma, C. Meessen, A.G. Mello, G. Merino, P. Morettini, E. Moyse,
A. Nairz, A. Negri, N. Nikitin, A. Nisati, C. Padilla, F. Parodi, V. Perez-Reale, J.L. Pinfold, P. Pinto, G. Polesello, Z. Qian,
S. Rajagopalan, S. Resconi, S. Rosati, D.A. Scannicchio, C. Schiavi, T. Schoerner-Sadenius, E. Segura, T. Shears, S. Sivoklokov,
M. Smizanska, R. Soluk, C. Stanescu, S. Tapprogge, F. Touchard, V. Vercesi, A. Watson, T. Wengler, P. Werner, F.J. Wickens,
W. Wiedenmann, S. Wheeler, M. Wielers, H. Zobernig